# Are optical and X-ray AGN mostly disjoint?

Yu-Xiao Dai,[1]* Glennys R. Farrar[1]
[1]*Center for Cosmology and Particle Physics, New York University, New York, NY 10003, USA*

**ABSTRACT**

The relationship between the populations of optically and X-ray selected Active Galactic Nuclei (AGN) has been unclear due to divergent results from different studies. Arnold et al. (2009) claim that X-ray AGN are almost entirely disjoint from optical AGN, while the Swift-BAT 70-month hard X-ray survey reported that 553 of their 711 X-ray AGN are optical. In this work, we set out to understand this difference by cross-checking between these studies and examining their sampling and AGN-selection criteria. We also re-analyze the X-ray and optical AGN in 16 groups and clusters reported by Arnold et al. using our own optical spectrum fitting techniques. We find that 6 of the 8 X-ray AGN in the Arnold et al. sample are also optical AGN, contrary to Arnold et al.'s report that only 1 of the 8 X-ray AGN is also an optical AGN, thereby falsifies their conclusion that optical and X-ray AGN are nearly disjoint sets.

## 1 INTRODUCTION

Active Galactic Nuclei (AGN) are highly energetic central regions of galaxies powered by accretion of matter onto supermassive black holes (SMBH). AGN produce emission covering a broad range of wavelengths, from radio to $\gamma$-rays. There are many open questions regarding the properties and classifications of AGN, arising from the many physical processes occurring and detailed structures hidden from view due to AGN's anisotropic radiation patterns and absorption by surrounding material. Different empirical identification methods are developed in different bands, providing distinct windows on the physics of AGN (Padovani et al. 2017).

An optical AGN is identified based on the galaxy's optical properties. Objects with broad H$\alpha$ FWHM ≥ 1000 km/s are Type 1 AGN. Type 2 AGN are identified through their narrow line flux ratios [O III]$\lambda$5007/H$\beta$ and [N II]$\lambda$6584/H$\alpha$ being in the appropriate region on the 'BPT Diagram' (Baldwin et al. 1981). A galaxy is classified as an X-ray AGN if the X-ray luminosity exceeds the expected contribution from other sources of X-ray emission such as emission from X-ray binaries, thermal emission from hot gas, and supernova remnants associated with substantial recent star formation. This is done using relations between X-ray luminosity, K-band luminosity, and star formation rate (Sivakoff et al. 2008).

Many distinctions in observed properties of AGN are dominated by orientation effects rather than real physics properties. According to the unified model of AGN, broad optical emission lines are emitted from the broad-line region (BLR) located close to the SMBH, outside of which a surrounding obscuring dusty torus prevents us from directly viewing the BLR. Illuminated by the X-ray emission from the central engine, interstellar gas in the narrow-line region (NLR), located above and below the plane of the torus, emits narrow emission lines (Antonucci 1993; Urry & Padovani 1995; Padovani et al. 2017). Type 2 AGN's optical spectra exhibit by definition only the narrow emission lines and lack broad lines, interpreted as resulting from the observer's line of sight to the central engine intersecting the putative torus. In the case of Type 1 AGN, the central engine is unobscured, giving the broad lines in the optical spectrum. X-ray emission from AGN primarily originates above the SMBH in a corona of hot electrons, where UV photons radiated by the accretion disk are inverse Compton scattered to X-ray energies as powerful as a few hundreds of keV. This intrinsic X-ray emission is believed to be tightly related to the accretion process almost universally amongst AGN (Lusso & Risaliti 2016). Therefore, we would expect a strong correlation between the optical emission lines and the X-ray emission if AGN are accreting and ionizing the NLR at a constant rate (Berney et al. 2015).

However, there are still many unresolved questions as regards the unified model and our knowledge of AGN. For example, we may expect most optical AGN to be detected at X-ray wavelengths since X-rays are very penetrating and can pierce through obscuring columns of $\lesssim 10^{24}$ cm$^{-2}$ (Digby-North et al. 2010). This is inconsistent with results from Shen et al. (2006), who find that in 140 galaxies in the group environment, there are 5 optical AGN while none of them are X-ray detected, and other studies report that only a small fraction of the optically selected AGN are detected by X-ray (Arnold et al. 2009; Digby-North et al. 2010). Starting from X-ray surveys, a similar disconnection between optically and X-ray selected AGN is also found. Some studies have shown that only a small fraction of the X-ray AGN associated with groups and clusters also possess optical AGN signatures (Martini et al. 2002, 2006; Davis et al. 2003; Finoguenov et al. 2004).

In this work, we seek to understand the apparently contradicting results that in Swift-BAT's 70-month hard X-ray survey (Baumgartner et al. 2013, hereafter B+13), 77.8% of X-ray AGN are associated with optical counterparts, while Arnold et al. (2009, hereafter A+09) reported that in the galaxies from their sample of 16 groups and clusters, they found 14 optical AGN and 8 X-ray AGN but only 1 is classified as both. It is evident that the fractions reported in B+13 and A+09 are incompatible, as can be seen from the following statistical analysis: Suppose we accept Swift-BAT's fraction of X-ray AGN that are also Type 2 AGN, i.e. 553/711, which we denote by $f^B_{o/X}$ = 0.778, the expected number of optical AGN in Arnold et al. (2009)'s 8 X-ray AGN ($N^A$) is $N^A \times f^B_{o/X}$ = 8×0.778 = 6.222. For a Poisson distribution with mean $\lambda$ = 6.222, we have the probability of observing one or none optical AGN is $P(X \leq 1)$ = 0.0143; This is unlikely enough to require further investigation. In this paper, we address of origin of this difference.

* E-mail: yd742@nyu.edu

In order to understand the disagreement between A+09 and B+13, we begin by reinvestigating the AGN identification methods used in Arnold et al.'s work in Section 2. In Section 3, we introduce the disagreement between studies on whether X-ray and optical AGN are disconnected, and then in Subsection 3.2, we use the method developed by Zaw, Chen & Farrar (2017, submitted) to analyze the spectra of the AGN reported by Arnold et al. Our conclusions are summarized in Section 4. In the appendix we include a correlation analysis between the [O III]$\lambda$5007 and hard X-ray luminosity of SDSS galaxies that are surveyed by both ZCF and Swift-BAT.

## 2 AGN IN ARNOLD ET AL.

Arnold et al. (2009) searched for optical and X-ray AGN in 10 groups and 6 clusters in the low-redshift range with 0.0203 < $z$ < 0.0770, picked from XMM-Newton observations and previous samples in SDSS DR6 (Adelman-McCarthy et al. 2008), 2.5m du Pont Telescope (Martini et al. 2006), and Chandra X-ray Observatory (Martini et al. 2006; Sivakoff et al. 2008). From this study, they report that X-ray AGN are nearly disjoint from the optical AGN. In the following paragraphs, we examine A+09's galaxies and AGN identification.

### 2.1 X-ray AGN in Arnold et al.

Arnold et al.'s X-ray AGN identification procedure can be summarized as follows. They combined images created by three XMM detectors in the 0.5-8 keV band using the SAS task EMOSAIC, and ran the task EWAVELET with a detection threshold of 5$\sigma$ to identify X-ray sources. These were then compared to the known members that are within 13' of the field center. Matches are determined if a known member is within 2" (the 1$\sigma$ positional uncertainty of XMM is 1-2"). They excluded central galaxies in their study for X-ray groups, poor clusters, and rich clusters due to difficulties caused by diffuse X-ray emission. For each member detected by XMM, they extracted a surface brightness profile and source spectra to determine the galaxy's X-ray emission.

Five main sources that can produce substantial ($L_X > 10^{40}$ erg s$^{-1}$) broad-band X-ray emission in the band 0.3-8 keV from galaxies are AGN, low-mass X-ray binaries (LMXBs), high-mass X-ray binaries (HMXBs), thermal emission from hot gas from the host galaxy or a galaxy cluster, and emission from the supernova remnants associated with substantial recent star formation (Fabbiano 2006; Boroson et al. 2011; Arnold et al. 2009). Relations between X-ray luminosity, K-band luminosity, and star formation rate were used by A+09 to determine the expected contribution from other processes than AGN and to classify a source as AGN if the X-ray luminosity exceeded the calculated contribution from the other emission sources (Sivakoff et al. 2008). A+09 considered two spectral models (1) a power-law component (LMXBs, AGN) (2) a thermal component (emission from hot gas). They constrained both components for galaxies with many counts, and only the power-law component for cases with fewer counts.

Out of the 14 X-ray sources, 8 have sufficient S/N to determine which model to fit: LMXBs, HMXBs, and an AGN component only (no thermal by spectra). Four sources cluster on the LMXB relations, and thus are considered inactive. The other 4 are at least 2$\sigma$ more X-ray luminous than LMXBs alone, and thus are considered X-ray AGN. Six have such faint X-ray emission that their X-ray spectra could not be accurately modeled. A+09 used LMXB + thermal emission from hot gas (modified (Sivakoff et al. 2008) soft band (0.5-2keV) results for broad-band (0.3-8keV)), assumed $\Gamma$ = 1.7, and calculated the X-ray luminosity. Five are at least 2$\sigma$ more X-ray luminous than LMXBs + thermal and are therefore identified as X-ray AGN. One was excluded because $L_X < 10^{41}$ erg s$^{-1}$. None of these 14 galaxies has sufficient star formation (estimated from H$\alpha$ flux (Kennicutt Jr 1998; Brinchmann et al. 2004)) to contribute significantly to the X-ray luminosity (calculated according to (Grimm et al. 2003)). Thus, they classified 8 of them as X-ray AGN.

### 2.2 Optical AGN in Arnold et al.

Taking data from the NORAS catalog of the MPA-JHU catalog of galaxies in SDSS Data Release 7[1], A+09 identified optical AGN based on the galaxy's [O III]$\lambda$5007/H$\beta$ and [N II]$\lambda$6584/H$\alpha$ line ratios on the 'BPT Diagram'. A galaxy is identified as an AGN if (a) its line flux measurements for [N II] and H$\alpha$ have S/N>3; (b) the two line ratios place it on the BPT diagram above the Kewley et al. (2001, hereafter Ke+01) line, which corresponds to the theoretical upper limit for pure starburst models so that a substantial AGN contribution to the line fluxes is required to move a galaxy above this line. Among the 349 galaxies in the MPA-JHU catalog investigated by A+09 using the MPA-JHU DR7 catalog, 116 have bright enough spectral lines to be placed on the BPT diagram, from which 14 are identified as optical AGN. Arnold et al. classified only one galaxy as both an X-ray AGN and an optical AGN.

## 3 THE DISAGREEMENT BETWEEN ARNOLD ET AL. & BAUMGARTNER ET AL.

Arnold et al. (2009) contend that AGN identified via emission-line diagnostics are 'nearly disjoint' from AGN that are X-ray-selected based on the result that there is only one galaxy that is classified both as an X-ray AGN and an optical AGN in their 10 groups and 6 clusters. If we look at the fraction of X-ray AGN that are also optical AGN, A+09 repored 1/8 (12.5%), considerably smaller than the 77.8% from the Swift-BAT 70-month survey.

A closer examination summarized in Table 1 & 2 indicates that A+09's argument can be improved in several ways.

(i) Checking the BPT status of the 8 X-ray AGN classified by A+09 using MPA-JHU DR7's fluxes in MPA-JHU's updated DR8

---

[1] http://www.mpa-garching.mpg.de/SDSS/DR7/

**Table 1.** 14 BPT AGN reported by Arnold et al. (2009). Galaxy properties for DR8 spectra from MPA-JHU are based on the methods described in Brinchmann et al. (2004). The galaxies which are in 2MRS (column 5) are checked for ZCF17 identifications (column 6). In the last two columns, we present the result of running the ZCF procedure on the spectra of all 14 galaxies ('Sy1': Type 1 AGN, 'Sy2': Type 2 AGN, 'Ke+01': Type 2 AGN by Kewley et al. (2001), 'Ka+03': Type 2 AGN by Kauffmann et al. (2003), 'Emi': Emission line galaxy but not AGN, blank: not emission line galaxy).

| O# | A+09 BPT AGN ID (MPA-JHU DR7) | MPA-JHU DR8 AGN | | 2MRS | ZCF17 | ZCF Method | |
|---|---|---|---|---|---|---|---|
| | | High S/N | Low S/N | | | Ke+01 | Ka+03 |
| 1 | 2MASXJ01095902+1358155 | | ✓ | | | Sy2 | Sy2 |
| 2 | SDSSJ010957.88+140320.1 | | ✓ | | | Sy2 | Sy2 |
| 3 | SDSSJ011021.57+135421.4 | | ✓ | | | Sy2 | Sy2 |
| 4 | 2MASXJ07462331+3101183 | ✓ | | | | Sy2 | Sy2 |
| 5 | 2MASXJ07470054+3058205 | | ✓ | | | Sy2 | Sy2 |
| 6 | 2MASXJ08445063+4302479 | ✓ | | ✓ | ✓ | Sy2 | Sy2 |
| 7 | SDSSJ100311.10+323511.3 | | ✓ | | | Sy2 | Sy2 |
| 8 | 2MASXJ10213991+3831195 | | ✓ | | | Sy2 | Sy2 |
| 9 | 2MASXJ11223691+6710171 | | | | | | |
| 10 | SDSSJ112425.38+671940.0 | | ✓ | | | Sy2 | Sy2 |
| 11 | 2MASXJ12041899+0150546 | ✓ | | ✓ | ✓ | Sy2 | Sy2 |
| 12 | 2MASXJ12230667+1037170 | ✓ | | ✓ | ✓ | Sy2 | Sy2 |
| 13 | 2MASXJ12225772+1032540 | | ✓ | ✓ | ✓ | Sy2 | Sy2 |
| 14 | 2MASXJ13241000+1358351 | ✓ | | ✓ | ✓ | Sy2 | Sy2 |

**Table 2.** 8 X-rays AGN reported by Arnold et al. (2009). Column 3 (labeled 'BPT?') gives Arnold et al. (2009)'s optical AGN classifications. Other columns are as in Table 1.

| X# | A+09 AGN | | MPA-JHU DR8 AGN | | 2MRS | ZCF17 | ZCF Method | |
|---|---|---|---|---|---|---|---|---|
| | X-ray ID | BPT? | High S/N | Low S/N | | | Ke+01 | Ka+03 |
| 1 | 2MASXJ07463295+3101213 | | | | | | Sy2 | Sy2 |
| 2 | 2MASXJ08445063+4302479 | ✓ | ✓ | | ✓ | ✓ | Sy2 | Sy2 |
| 3 | 2MASXJ10230356+3838176 | | | | | | Sy1 | Sy2 |
| 4 | 2MASXJ10223745+3834447 | | ✓ | | ✓ | | Emi | Sy2 |
| 5 | 2MASXJ10220069+3829145 | | | | | | Sy1 | Sy1 |
| 6 | 2MASXJ11231618+6706308 | | | | | | | |
| 7 | 2MASXJ11221610+6711219 | | | | | | Sy2 | Sy2 |
| 8 | SDSSJ112333.56+671109.9 | | | | | | Sy2 | Sy2 |

catalog[2], we discover in addition to the galaxy that A+09 reported (X#2 in Table 2), another one (X#4) also satisfies the BPT AGN criteria, increasing the fraction of BPT AGN.

(ii) One galaxy (O#9) that is classified as an AGN by MPA-JHU's DR7 is absent in either high or low S/N AGN category by the MPA-JHU DR8 release[3].

(iii) What A+09 refers to as AGN by emission-line classification or optical AGN are, as a matter of fact, *Type 2 AGN only*, leaving out the entire Type 1 AGN population. According to Swift-BAT, the fraction of Type 2 AGN in X-ray AGN is 261/711 = 36.7%, which is closer to A+09's Type 2 AGN fraction. This clarification significantly alleviates the discrepancy.

### 3.1 Zaw, Chen & Farrar All-sky optical AGN catalog

Apart from the aforementioned problems in A+09's analysis, Chen, Zaw & Farrar (2017, submitted) have recently shown that the very sample A+09 relied on — the MPA-JHU catalog — misclassified 40% of low luminosity AGN, so we re-examine the classification with the improved method described in Zaw, Chen & Farrar (2017, submitted, ZCF17 hereafter). ZCF17 introduced the first all-sky optical AGN catalog, which uses the Two Micron All-Sky Survey's Redshift Survey (Huchra et al. 2012, 2MRS) as a parent sample. AGN are identified based on optical spectra obtained from SDSS, 6dF, FAST, and CTIO, representing ~80% of the 2MRS spectra. Objects with broad H$\alpha$ FWHM $\geq$ 1000 km/s are identified as Type 1 AGN. Type 2 AGN are identified through placing their narrow line flux ratios [O III]$\lambda$5007/H$\beta$ and [N II]$\lambda$6584/H$\alpha$ on the BPT Diagram and further requiring S/N > 1.64 for at least 3 of the emission lines.

We searched for A+09's 14 BPT AGN in 2MRS and in its subset the ZCF17 optical AGN catalog and found that for 5 of the A+09 BPT AGN that are in 2MRS, all are also identified ZCF AGN. The status of each of these BPT AGN in various surveys described in this section is summarized in Table 1. We then searched for A+09's X-ray AGN, two of which are AGN according to MPA-JHU's DR8 catalog as we mentioned in the beginning of the section. These two happen to be the only galaxies that are surveyed by 2MRS, ergo analyzed by ZCF17, but only one (X#2) is identified as a Type 2 AGN in ZCF17. Up to this point, we have found an inconsistency between the SDSS MPA-JHU catalog and the ZCF17 all-sky optical

---

[2] The latest MPA-JHU catalog (http://www.sdss.org/dr12/spectro/galaxy_mpajhu/) classifies SDSS Data Release 8 galaxies into subsamples based on their emission line properties using the methodology described in (Brinchmann et al. 2004). The AGN population consists of the galaxies above the Ke+01 line in the BPT diagram (S/N > 3 for [N II], H$\alpha$, [O III] and H$\beta$). Low S/N AGN, or 'low S/N LINER' as on MPA-JHU's website and in (Brinchmann et al. 2004), are galaxies that have [N II]/H$\alpha$ > 0.6 (S/N > 3 for both lines) while [O III] or H$\beta$ has S/N < 3.
[3] According to MPA-JHU's DR8 catalog, 5 of the 14 BPT AGN reported by Arnold et al. (2009) are classified as AGN, with the other 8 belonging to low S/N AGN.

AGN catalog, in that the galaxy X#4 is present in both SDSS DR7/8 and 2MRS, meaning it is analyzed by both MPA-JHU and ZCF17, but the former classifies the galaxy as an AGN while the latter does not. This is consistent with the MPA-JHU 40% mis-identified rate for low luminosity AGN pointed out by Chen, Zaw & Farrar (2017, submitted).

### 3.2 Analysis of Arnold et al.'s AGN using the spectrum-fitting method by Zaw, Chen & Farrar

We examined A+09 X-ray and BPT AGN using the spectral-fitting method developed by Chen, Zaw & Farrar (2017, submitted) that was employed in creating the ZCF17 catalog to clarify these galaxies' AGN candidacy and make a conclusive judgment on the 'discrepancy' between A+09 and Swift-BAT. In this method, emission lines are isolated by subtracting the stellar absorption and continuum emission component from the spectrum. The host-galaxy contribution of the spectrum is fit using the full-spectrum-fitting program pPXF (Cappellari & Emsellem 2004) as a linear combination of the single stellar population (SSP) templates from the MILES stellar library[4], which consists of 1000 stars with a wavelength range of 3525-7500Å and a spectral resolution of 2.5Å. ZCF reprocessed SDSS spectra to cross check with published line fluxes by SDSS teams, and found that the line ratios measured by using MILES templates are systematically lower than the SDSS published values obtained using the Bruzual & Charlot (2003, BC03) templates. Analyses show the discrepancies are due to the known problems in BC03 (Chen, Zaw & Farrar 2017, submitted).

The results we get from running the ZCF analysis code on the AGN reported in A+09 are summarized in the last two columns in Table 1 and Table 2. Of the 14 BPT AGN in A+09, 13 are flagged as either AGN or low S/N AGN by MPA-JHU DR8. These 13 galaxies are also identified as Type 2 AGN by our examination (satisfying both Ke+01 and Ka+03 criteria), agreeing to MPA-JHU DR8's AGN classification. For A+09's 8 X-ray AGN, we identify 6 of them as optical AGN using the ZCF method, with two Type 1 and four Type 2 AGN. However, as we mentioned in Section 2.1, MPA-JHU found only one Type 2 in DR7 and two Type 2 in DR8, missing most of the optical AGN amongst them. Therefore, the fraction of optical AGN (Type 1 and 2 combined) in A+09's X-ray AGN by our spectral analysis is 75.0%, significantly higher than the 12.5% originally reported in Arnold et al. (2009).

## 4 SUMMARY AND CONCLUSION

The study reported here was prompted by the claim of Arnold et al. (2009) that optical and X-ray AGN in groups and clusters form nearly disjoint sets. This is at variance with Baumgartner et al. (2013) who examined a much larger data set from the Swift-BAT 70-Month survey. In that study, 77.8% of the X-ray AGN are associated with counterparts that are optical AGN. A puzzle that was raised is why the portion of optical AGN in X-ray AGN is much greater according to Swift-BAT. One possibility advanced by Arnold et al. is the different selection criteria of the two studies: Swift-BAT selected on the basis of X-ray flux and Arnold et al. selected based on membership in groups and clusters.

In this work, we examined Arnold et al.'s source of emission line fluxes and errors, which is MPA-JHU DR7, and Arnold et al.'s classification criteria. First, we found, using the most recent MPA-JHU DR8's classification, that one of the 14 optical AGN classified as optical AGN by DR7 is not by DR8, and that there is an additional X-ray AGN identified as optical AGN by DR8. We then uncovered that optical-selected AGN in Arnold et al. meant Type 2 AGN only, leaving out the entire Type 1 population. After including both Type 1 and 2 in the analysis, the fraction of X-ray AGN that are also optical AGN increases greatly. Finally, we investigated all the AGN reported in Arnold et al. using the identification procedure developed by Zaw, Chen & Farrar (2017, submitted). We discovered that Arnold et al., relying on the MPA-JHU catalog of BPT AGN, failed to correctly identify that 5 of the 8 X-ray AGN are also optical AGN. This resolves the puzzle of apparent disagreement with Swift-BAT.

## ACKNOWLEDGEMENTS

We are grateful to Yanping Chen for allowing us to use her code to perform the 'ZCF' AGN classifications shown in Tables 1 and 2, and for her help in using it. We also acknowledge helpful discussions with Yanping Chen and Ingyin Zaw on topics related to this analysis, and G.R. Farrar acknowledges with thanks her collaboration with them on the general problem of proper classification of optical AGNs.

The research of G.R. Farrar has been supported in part by the U.S. National Science Foundation (NSF), Grants NSF-1212538 and NSF-1517319.

---

[4] http://www.iac.es/proyecto/miles/pages/webtools/tune-ssp-models.php

## APPENDIX A: THE RELATIONSHIP BETWEEN AGN'S X-RAY LUMINOSITY AND ITS CANDIDACY FOR BEING AN OPTICAL AGN

The disagreement between Arnold et al. (2009) and Baumgartner et al. (2013) could be explained if a larger portion of high-luminosity X-ray AGN tend to satisfy optical AGN classification criteria than those that are less luminous in X-ray, since the luminosity of X-ray AGN in A+09 is ∼2 dex lower than those in B+13. This hypothesis later turned out to be unnecessary since when we properly determined whether A+09's AGN are really AGN through analyzing their spectra directly, the discrepancy disappeared. However, our investigation of this hypothesis did produce results that could be useful to further studies on AGN's properties.

To examine this hypothesis, we investigate the relationship between X-ray and optical AGN by comparing the hard X-ray (14-195 keV) and [O III] luminosities. The results are presented in Fig. A. The 121 Swift-BAT sources that have counterparts in the SDSS subsample[5] of ZCF are represented by red dots (Type 1) and blue triangles (Type 2). Contrary to the previous study by Heckman et al. (2005) who analyzed 47 local AGN from the RXTE slew survey and found a correlation between their [O III] luminosity and 3-20 keV X-ray luminosity, we do not find a significant correlation, in agreement with Berney et al. (2015). The lack of correlation means AGN's optical emission strength is not related to its X-ray luminosity, rendering the hypothesis unlikely. This is consistent with the findings of our spectroscopic analysis described in Section 3.2.

We also include the X-ray and optical AGN of A+09 in Fig. A as arrows on axes for those with [O III] flux (from running the ZCF procedure) only, and as green circled points for those with both X-ray and [O III] fluxes. Fluxes and luminosities of X-ray AGN in A+09 are measured in the 0.3-8 keV band. By approximating this with the 2-10 keV band and taking 4 AGN with multi-band fluxes in Fioretti et al.

[5] Only the SDSS galaxies had absolute calibrated spectra.

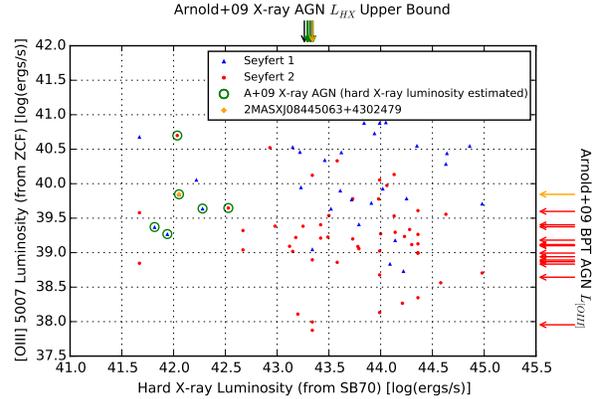

**Figure A1.** [O III]$\lambda$5007 luminosity in [log(ergs/s)] (from ZCF17) vs. hard X-ray luminosity in [log(ergs/s)] (from Swift-BAT). Data points with no circles around are Swift-BAT sources that have ZCF counterparts, with red dots representing Type 1 AGN and blue triangles representing Type 2. A+09's AGN are placed in this plot using arrows. Arrows pointing to the left are [O III]$\lambda$5007 luminosities of 13 BPT AGN. Those pointing downwards are the maximum possible X-ray luminosity for A+09's AGN in the Swift-BAT band derived from their redshift and the minimum flux of the Swift-BAT 70-month survey sources. Among them, four Type 2 and two Type 1 AGN are identified by the ZCF procedure, so we have their [O III]$\lambda$5007 luminosity values. They are placed on the scatter plot denoted by dots and trangles with green circles around. 2MASXJ08445063+4302479, the single galaxy identified by Arnold et al. (2009) as both an X-ray and optical AGN, is separately indicated by a circled orange diamond.

(2013) as a reference, we estimate the X-ray luminosity in the 14-195 keV band of these 8 X-ray AGN (arrows pointing downwards) to be in the range 41.82-42.69 ± 0.22 [log(ergs/s)] with an estimated flux upper limit at $8.28 \times 10^{13}$ erg/s/cm$^2$. This is more than 1 dex under the minimum flux requirement of the Swift-BAT X-ray AGN survey reported in B+13. The estimated luminosity X-ray luminosity of these AGN is also much lower than most of the Swift-BAT AGN. The arrows above represent the maximum possible X-ray luminosity for A+09's AGN in the Swift-BAT band derived from their redshift and the minimum flux of the Swift-BAT 70-month survey sources. The [O III] luminosity of these low-hard-X-ray-luminosity AGN is at the same level as the much more luminous ones in Swift-BAT, in accordance with previous studies such as (Winter et al. 2010) and (Berney et al. 2015), and disagrees with the strong correlation reported in (Heckman et al. 2005).